\documentclass[aps,twocolumn,epsf,epsfig,showpacs]{revtex4}
\usepackage{graphicx}
\usepackage{color}
\begin{document}
\newcommand{\be}{\begin{equation}}
\newcommand{\ee}{\end{equation}}
\newcommand{\beqa}{\begin{eqnarray}}
\newcommand{\eeqa}{\end{eqnarray}}
\def\nn{\nonumber}
\def\l({\left(}
\def\r){\right)}
\def\gr{$\gamma$-ray}

\title{Gamma-ray induced cascades and magnetic fields in intergalactic medium}
\author{A.~Elyiv}
\affiliation{Main Astronomical Observatory National Academy of Sciences of Ukraine, 27 Akademika Zabolotnoho St.
03680 Kyiv, Ukraine}
\author{A.~Neronov}
\affiliation{ISDC Data Center for Astrophysics, Chemin d'Ecogia 16, 1290 Versoix,
Switzerland and Geneva Observatory, 51 ch. des Maillettes, CH-1290 Sauverny,
Switzerland}
\author{D.V.~Semikoz}
\affiliation{APC, 10 rue Alice Domon et Leonie Duquet, F-75205 Paris Cedex 13, France}
\affiliation{Institute for Nuclear Research RAS, 60th October Anniversary prosp. 7a,
Moscow, 117312, Russia}

\begin{abstract} 
We present the results of Monte-Carlo simulations of three-dimensional electromagnetic cascade 
 initiated by interactions of the multi-TeV \gr s with the cosmological infrared/optical photon background in the intergalactic medium. Secondary electrons in the cascade are deflected by the intergalactic magnetic fields before they scatter on CMB photons.  This leads to extended 0.1-10 degree scale emission at multi-GeV and TeV energies around extragalactic sources of very-high-energy \gr s. The morphology of the extended emission depends, in general, on the properties of magnetic fields in the intergalactic medium. Using Monte-Carlo simulated  data sets, we demonstrate that the decrease of the size of extended source with the increase of energy allows  to measure weak magnetic fields  with magnitudes in the range from $\le 10^{-16}$ G to $10^{-12}$G if they exist in the voids of the Large Scale Structure.  
\end{abstract}
\maketitle

%%%%%%%%%%%%%%%%%%%%%%%%%%%%%%%%%%%%%%%%%%%%%%%%%%%
\section{Introduction}
%%%%%%%%%%%%%%%%%%%%%%%%%%%%%%%%%%%%%%%%%%%%%%%%%%%

 Spectral and timing properties of astronomical sources of very high-energy \gr s could be strongly affected by the development of electromagnetic cascades on the way from the source to the Earth. These cascades could be initiated by interactions of the \gr s with the ambient radiation fields inside the \gr\ source, in the source host galaxy and in the Milky Way galaxy, as well as with cosmological photon fields in the intergalactic space. The ubiquity of particle cascades has two-fold consequences. On one, pessimistic, side, they complicate the interpretation of the observational data in the very-high-energy (VHE, 0.1-10~TeV) $\gamma$-ray band. On the other, optimistic, side, with enough spectral and angular resolution, one can quantify the influence of the cascade on the observed source signal and not only reconstruct the spectrum of the primary $\gamma$-ray source, but also use the information about properties of the cascade to study the physical characteristics of the medium in which the cascade has developed.  

An illustration of such possibility is given by the observations of the effect of absorption of VHE $\gamma$-rays on the extragalactic background light (EBL). Observations of this effect in the spectra of distant blazars are now routinely used to constrain the EBL spectrum and, in this way, the models of cosmological evolution of galaxies and stellar populations  (see  \cite{aharonian07a} for a recent review). 

In the derivation of the constraints on the EBL from the observations of absorption of VHE $\gamma$-ray flux from distant blazars one usually adopts an assumption that secondary emission from the $e^+e^-$ pairs, deposited in the intergalactic space by the absorbed $\gamma$-rays, is not detectable. This assumption is valid if the trajectories of the secondary pairs are significantly deflected by the magnetic fields during their radiative cooling. This is true if the strength of the extragalactic magnetic fields (EGMF) is higher than  $\sim 10^{-12}$~G \cite{Aharonian:halo,neronov07b}.  

However, the strength of EGMF is not measured (see \cite{beck08} for a review of measurements of cosmic magnetic fields), so that there is no direct way to check the validity of the the assumption of the non-detection of the cascade emission. It is also not possible to constrain the values of EGMF from the cosmological models of the origin of magnetic fields in the Universe. 

The problem of the origin of $1-10\ \mu$G magnetic fields in galaxies  and galaxy clusters is one of the long standing problems of astrophysics/cosmology~(see \cite{kronberg,primordial_review,primordial_review2} for reviews).  On the galaxy scales, such fields are thought to be produced  via dynamo mechanism \cite{dynamo,dynamo2} from "seed" primordial fields of unknown origin. Recent observations of the presence of strong, $\sim 10\ \mu$G, magnetic fields in the high-redshift galaxies \cite{bernet,wolfe} point to the fact that the dynamo mechanism is extremely efficient and/or the seed magnetic fields are quite strong  so that the 10~$\mu$G magnetic fields are generated already at large redshifts. 

The dynamo mechanism could work also on the scales of galaxy clusters \cite{tribble93}, although the seed magnetic fields would be amplified by a much smaller factor in this case. The fact that the observed strength of magnetic field in galaxy clusters is comparable to that of the individual galaxies \cite{clusters} implies that the seed fields in the galaxy clusters should be orders of magnitude higher than the seed magnetic fields for the galactic dynamos. Recent simulations show that such stronger seed magnetic fields could be produced via magnetized winds from the cluster galaxies \cite{bertone,donnert}.

The origin of the seed magnetic fields for the galactic dynamos is largely uncertain.  All the models of production of such seed fields invoke the difference of mobility of electrons and protons, which results in production of electric currents and of the "battery-like" effects at different stages of cosmological evolution, from the phase transitions in the early Universe, to the formation of first galaxies \cite{primordial_review,primordial_review2}. A common feature of all the theoretical models is a prediction that a larger or smaller fraction of the space outside galaxies, galaxy clusters and filaments of the large scale structure has to be filled with very weak magnetic field, with the strength approaching that  of the primordial seed fields. Depending on the model, the predictions of the typical strength of the EGMF in the voids of the large scale structure range from $\le 10^{-20}$~G  up to $\sim 10^{-9}$~G (see e.g. \cite{primordial_review,primordial_review2,defl_Sigl,defl_Dolag,bruggen}).   Moderate observational limit on
the present day strength of EGMF $B<10^{-9}$G comes from the limit on rotation measure of emission from distant quasars~\cite{primordial_review}. Similar restriction comes from the analysis of anisotropies 
of  the cosmic microwave background~\cite{Barrow:1997mj,blasi}.

Gamma-ray observations could, in principle, be used to constrain properties of the EGMF using the imaging  \cite{neronov07b} and/or timing \cite{plaga,Murase:2008pe} of the \gr\ signal.  The idea  is to observe or, at least, constrain the properties of the inverse Compton (IC) emission produced by the $e^+e^-$ pairs deposited by the absorbed \gr s in the intergalactic space. If the EGMF strength is below $10^{-12}$~G (plausible assumption, in the view of the cosmological models of the origin of magnetic fields), one expects that the deflections of the secondary electrons and positrons by EGMF should result in the appearance of an EGMF-dependent extended emission around initially point sources and/or in an EGMF-dependent time delay of the emission from the secondary pairs.  If the EGMF is stronger than  $10^{-12}$~G, VHE $\gamma$-ray observations still can be used to derive a {\it lower} limit on the EGMF strength, because strong EGMF $\gg 10^{-12}$~G can still be revealed via the extended emission around VHE sources \cite{Aharonian:halo} with properties not directly dependent on the EGMF structure. 

In what follows we explore the possibility to measure the EGMF using the data of VHE \gr\ observations of extragalactic sources. In the Section \ref{sec:analytic} we extend ideas of ref.~\cite{neronov07b} on developement of electromagnetic cascade in the intergalactic medium and find the range of EGMF strengths which lead to observable extended emission around extragalactic \gr\  point sources at GeV and TeV energies. In the Section  \ref{sec:numeric} we model the development of the \gr\ induced cascades in the intergalactic medium via Monte-Carlo simulations. This allows us to simulate the cascade-induced extended emission around initially point \gr\ sources and to study the dependence of the properties of this extended emission on the strength of EGMF. Using the simulated data sets, we demonstrate that the assumed strength of EGMF can be calculated from the measurement of the energy dependence of the size of the extended emission. We summarize our results in the Section \ref{sec:conc}.

%%%%%%%%%%%%%%%%%%%%%%%%%%%%%%%%%%%%%%%%%%%%%%%%%%%
\section{Electromagnetic cascade in intergalactic medium}
\label{sec:analytic}
%%%%%%%%%%%%%%%%%%%%%%%%%%%%%%%%%%%%%%%%%%%%%%%%%%%

Multi-TeV $\gamma$-rays emitted by distant point sources are not able to propagate over large distances because of the absorption in interactions with EBL. The pair production on EBL reduces the flux of  $\gamma $-rays from the source by
\begin{equation}
\label{absorb}
F(E_{\gamma_0})=F_0(E_{\gamma_0})\exp\left[-\tau(E_{\gamma_0},z)\right],
\end{equation}
where $F(E_{\gamma_0})$ is the detected spectrum, $F_0(E_{\gamma_0})$ is the initial spectrum of the source and $\tau(E_{\gamma_0},z)$ is the optical depth with respect to the pair production on EBL, which is a function of the primary photon energy $E_{\gamma_0}$ and of the redshift of the source $z$. 

The $e^+e^-$ pairs of the energy $E_e$ produced in interactions of multi-TeV $\gamma$-rays with EBL photons produce secondary $\gamma$-rays via IC scattering of the Cosmic Microwave Background (CMB) photons to the energies
\begin{equation}
E_{\gamma}\simeq \epsilon_{CMB}\frac{E_e^2}{m_e^2}\simeq 6
\left[\frac{E_{\gamma_0}}{100 \mbox{ TeV}}\right]^2\mbox{ TeV}
\label{Esec}
\end{equation}
where $\epsilon_{CMB}=6\times 10^{-4}$~eV is the typical energy of CMB photons. In the above equation we have assumed that the energy of primary $\gamma$-ray is $E_{\gamma_0}\simeq 2E_e$. Upscattering of the infrared/optical background photons gives sub-dominant contribution to the IC scattering spectrum because the energy density of CMB is much higher than the density of the infrared/optical background.

 Deflections of $e^+e^-$ pairs produced by the $\gamma$-rays, which were initially emitted slightly
away from the observer, could lead to "redirection" of the secondary cascade photons toward the observer. This effect leads to the appearance of extended emission around an initially point source of \gr s. 

The cascade electrons loose their energy via IC scattering of the CMB photons within the distance
\begin{equation}
D_e=\frac{3m_e^2c^3}{4\sigma_TU_{\rm CMB}E}\simeq 2.2\times 10^{22}\left[\frac{E_e}{50\mbox{ TeV}}\right]^{-1}\mbox{ cm}
\end{equation}
Comparing this to the Larmor radius in the magnetic field $B$,
\begin{equation}
R_L=\frac{E_e}{eB}\simeq 1.7\times 10^{24}\left[\frac{B}{10^{-13}\mbox{ G}}\right]^{-1}\left[\frac{E_e}{50\mbox{ TeV}}\right]\mbox{ cm}
\end{equation}
one can find typical deflection angle of the cascade photons
\begin{equation}
\Delta=\frac{D_e}{R_L}\simeq 0.7^\circ\left[\frac{B}{10^{-13}\mbox{ G}}\right]\left[\frac{E_e}{50\mbox{ TeV}}\right]^{-2}
\end{equation}
A simple geometrical calculation \cite{neronov07b} shows that the the bulk of the IC upscattered cascade \gr s of the energy $E_\gamma$ arrives within an angle
\begin{eqnarray}
\Theta_{\rm ext}&=&\frac{D_{\gamma_0}}{D}\Delta\simeq \frac{0.7^\circ}{\tau(E_{\gamma_0},z)}
\left[\frac{E_\gamma}{6\mbox{ TeV}}\right]^{-1}\left[\frac{B}{10^{-13}\mbox{ G}}\right]
\label{Thetaext}
\end{eqnarray}
where $D$ is the distance to the source and $D_{\gamma_0}$ is the mean free path of the primary \gr s. The above expression is valid for $\tau(E_{\gamma_0},z)>1$. If the extension of the cascade-induced "glow" around initially point source is determined by the deflections of the cascade electrons by the EGMF, the extension of the "glow", $\Theta_{\rm ext}$ is expected to shrink inversely proportional to the \gr\ energy. 
 
If $\tau(E_{\gamma_0},z)>1$, most of the power of the primary photon beam at the energy $E_{\gamma_0}$ is transfered to the power of electromagnetic cascade, so that the the total flux of the extended source is of the order of the primary source flux at the energy $E_{\gamma_0}$.  The extended source flux at a given energy $E_\gamma$ is further modified by the absorption on the way toward the Earth, so that the resulting flux of the extended source at the energy $E_\gamma$ is 
$F_{\rm ext}(E_\gamma,\Theta_{\rm ext})\simeq \left(e^{\tau(E_{\gamma_0},z)}-1\right)e^{-\tau(E_\gamma,z)}F(E_{\gamma_0})$.

The above estimate of the flux of the extended source is obtained under assumption that the magnetic field of the strength $B$ fills most of the intergalactic medium between the source and observer on the Earth. On the other hand, the line of sight can be intersected by the elements of the large scale structure, in which the magnetic field is much stronger than the typical EGMF. Strong deflections of electrons in the parts of the cascade traversing the large scale structure could lead to the suppression of the secondary cascade flux in the direction of observer. To take into account this effect, it is convenient to introduce the "volume filling factor", ${\cal V}(B)$, of the field of particular strength  $B$. The fraction of the path along the line of sight, occupied by the field of the strength $B$ is $\left[{\cal V}(B)\right]^{1/3}$ and the 
\begin{equation}
F_{\rm ext}(E_\gamma,\Theta_{\rm ext})\simeq \left[{\cal V}(B)\right]^{1/3}\frac{\left(e^{\tau(E_{\gamma_0},z)}-1\right)}{e^{\tau(E_\gamma,z)}}F(E_{\gamma_0})
\label{sext2}
\end{equation}

Combining Eqs. (\ref{Thetaext}), (\ref{sext2}), one can see that a measurement of the total flux and of the energy-dependent size of the cascade emission around an extragalactic point source enables to extract the information about the parameters of extragalactic magnetic field, $B$, ${\cal V}(B)$, along the line of sight. 

It is easier to detect an extended emission around a given point source if the angular size of the extended source is not too large. The energy of the detected \gr s, $E_\gamma$, and the EGMF, $B$, enter in the Eq. (\ref{Thetaext}) for the source size through a combination $E_\gamma/B$.  This means that stronger magnetic fields could be measured by the instruments sensitive at higher energies.  At the highest energies, the sensitivity of a telescope is determined mostly by the effective collection area $A_{\rm eff}(E_\gamma)$. Let us consider, as a reference, a source at the redshift  $z\simeq 0.03$, producing a flux $F(E_{\gamma_0})\sim 10^{-12}$~erg/cm$^2$s. If $E_{\gamma_0}\sim 100$~TeV, the optical depth $\tau(E_{\gamma_0},z)\sim 10$, so that all the primary \gr\ power is transfered to the extended source emission at the energy $E_\gamma\simeq 6$~TeV. The statistics of the extended emission signal is
\begin{eqnarray}
N_\gamma&\simeq& 2\times 10^3\left[\frac{F(E_{\gamma_0})}{10^{-12}\mbox{ erg/cm}^2\mbox{s}}\right]\left[\frac{A_{\rm eff}(E_\gamma)}{10\mbox{ km}^2}\right]\nonumber\\ &&\left[\frac{T_{\rm exp}}{50\mbox{ hr}}\right]\left[\frac{E_\gamma}{6\mbox{ TeV}}\right]^{-1}
\label{n_extend}
\end{eqnarray}
The statistics sufficient for the extended source analysis is thus achieved within the exposure time $T_{\rm exp}\sim 50$~hr (typical for the current generation ground-based \gr\ telescopes) with a telescope with an  effective area $A_{\rm eff}\sim 10$~km$^2$. Such effective area is achievable with the planned next-generation Cherenkov telescope array, optimized for the observations in the 10~TeV energy band \cite{plyashechnikov}. Otherwise, a comparable signal statistics is achievable with a much smaller collection area, $A_{\rm eff}\sim 0.1$~km$^2$, provided that the exposure time is $T_{\rm exp}\sim 1$~yr. This is achievable with the next-generation HAWC water Cherenkov telescope \cite{hawc}.

Extremely weak magnetic fields $B\le 10^{-16}$~G could be detectable through the degree-scale extended emission in the GeV, rather than TeV energy band, see Eq.(\ref{Thetaext}). The primary photons which lead to production of extended emission in the GeV band have energies $E_{\gamma_0} \sim 1-3$~TeV, see Eq.~(\ref{Esec}). The mean free path of such photons is much larger than that of the $\sim 100$~TeV photons,  so that detectable extended emission is expected around distant ($z\gg 0.1$), rather than nearby, sources in this case. Even smaller fields  can be probed with timing analysis of observed signal, as was discussed in the ref.~\cite{Murase:2008pe}. 

The extended emission in the GeV energy band has to be detected on top of the diffuse \gr\ background (DGRB). The characteristics of this background are uncertain at present. Previous measurements done by EGRET~\cite{egret,egret_new} indicate that a significant part of DGRB can be due to unresolved point sources (e.g. blazars). The true DGRB should contain at least contribution from cosmic ray interactions with CMB, but this contribution can be significantly below the EGRET value~\cite{diffuse_CR}.  In near future {\it Fermi} \cite{Fermi} will provide a new measurement of DGRB. Assuming that the intensity of the DGRB measured by {\it Fermi} will be by a factor $f_{Fermi}<1 $ lower than the EGRET measurement, we can estimate the sensitivity of {\it Fermi} for the detection of extended emission around high-redshift sources in the GeV energy band. The number of DGRB \gr s within a $\theta\sim 1$ degree region on the sky is 
\begin{equation}
N_{b} \simeq    10^2f_{Fermi} \left[\frac{E_\gamma}{1\mbox{ GeV}}\right]^{-1}  \left[\frac{A_{\rm eff}}{1\mbox{ m}^2}\right]\left[\frac{T_{\rm exp}}{1\mbox{ yr}}\right]  \left[\frac{\theta}{1^\circ}\right]^2
\label{n_diffuse}
\end{equation}
Re-scaling the Eq.~(\ref{n_extend}) for a brighter blazar $F(E_{\gamma_0})\sim 10^{-11}$~erg/cm$^2$s one finds that the signal statistics is  in this case:
\begin{eqnarray}
N_\gamma&\simeq& 2\times 10^3 \left[\frac{F(E_{\gamma_0})}{10^{-11}\mbox{ erg/cm}^2\mbox{s}}\right]\left[\frac{A_{\rm eff}(E_\gamma)}{1\mbox{ m}^2}\right]\nonumber\\ &&\left[\frac{T_{\rm exp}}{1\mbox{ yr}}\right]\left[\frac{E_\gamma}{1\mbox{ GeV}}\right]^{-1}
\label{n_extend_GeV}
\end{eqnarray}
Assuming the exposure time $T_{\rm exp}\sim 1$~yr, one finds that a moderate collection area $A_{\rm eff}\le 1$~m$^2$, typical for the space-born \gr\ telescopes, like {\it Fermi}, is sufficient for the study of expected extended emission.

Assuming that the sensitivity of a telescope is sufficient for the detection of extended emission, one can estimate the maximal measurable magnetic field from the condition that the source extension $\Theta_{\rm ext}(E_\gamma)$ does not exceed the size of the telescope's Field of View (FoV) $\Theta_{\rm FoV}$. Taking the source size at the energy $E_\gamma\simeq 1$~TeV as a reference, and substituting $\Theta_{\rm FoV}$ at the place of $\Theta_{\rm ext}$ in Eq. (\ref{Thetaext}), one can find that the maximal measurable field is
\begin{equation}
B_{\rm max}\simeq 3\times 10^{-13}\left[\frac{\Theta_{\rm FoV}}{1.5^\circ}\right]\left[\frac{E_\gamma}{1\mbox{ TeV}}\right]\left[\frac{\tau(E_{\gamma_0},z)}{10}\right]\mbox{ G}
\label{Bmax}
\end{equation}
An increase of the size of the FoV, expected with next-generation ground based \gr\ telescopes, such as the Cherenkov telescope array optimized for the 10~TeV energy band \cite{plyashechnikov}, or the next-generation wide field Cherenkov telescope array AGIS \cite{agis}, will provide, apart from a better sky coverage, an improvement of sensitivity for the detection of stronger EGMF. 

The weakest magnetic fields which can be probed is found from observation that $\Theta_{\rm ext}$ can not be measured if it becomes smaller than the size of the point spread function of the telescope, $\theta_{\rm PSF}$. Since $\Theta_{\rm ext}\sim E_\gamma^{-1}$ (see (\ref{Thetaext})), the largest source extension is achieved at lowest energies. Assuming $\theta_{\rm PSF}\sim 0.1^\circ$ (typical for the Cherenkov telescope arrays and for {\it Fermi}), one finds from (\ref{Thetaext})
\begin{equation}
B_{\rm min}\simeq 10^{-16} \left[\frac{\tau(E_{\gamma_0},z)}{1}\right] \left[\frac{E_\gamma}{0.1\mbox{ TeV}}\right]\left[\frac{\theta_{\rm PSF}}{0.1^\circ}\right]\mbox{ G}
\label{Bmin}
\end{equation}
The characteristics of a set-up optimized for the detection of weakest EGMF $B\sim 10^{-16}$~G, should be quite different from the one optimized for the detection of stronger EGMF, $B\sim 10^{-12}$~G. In this case, high sensitivity at low energies $E_\gamma\le 0.1$~TeV and good angular resolution are required. These can be achieved with the planned next-generation low-energy extensions of the ground-based Cherenkov telescope arrays, like HESS-II \cite{hess-2}, MAGIC-II \cite{magic-2} and CTA \cite{cta} or with {\it Fermi} \cite{Fermi}. Another important difference between the approaches for detection of strong and weak EGMF is different choice of sources. Strong EGMF could be detected via a study of extended emission around nearby ($z<0.1$) extragalactic TeV sources, while weak EGMF could be detected via studies of extended emission around sources of sub-TeV \gr s in the distant Universe ($z>0.1$).

%%%%%%%%%%%%%%%%%%%%%%%%%%%%%%%%%%%%%%%%%%%%%%%%%
\section{Numerical simulations}
\label{sec:numeric}
%%%%%%%%%%%%%%%%%%%%%%%%%%%%%%%%%%%%%%%%%%%%%%%%%

The EGMF-determined energy dependence of the size of the extended emission in the TeV energy band, given by Eq. (\ref{Thetaext}), provides a possibility to extract the information on the properties of the EGMF from the observational data. In the qualitative discussion of the previous section we have assumed, for the sake of simplicity, that each primary \gr\ produces electron and positron of energies equal to one-half of the primary \gr\ energy, $E_e=E_\gamma/2$, and that electron and positron emit IC photons of energies equal to $(E_e/m_e)^2\epsilon_{\rm CMB}$. In other words, we have adopted a "monochromatic approximation" for the production spectra of $e^+e^-$ pairs and of the IC scattered photons. A scatter of the energies of cascade particles can, in principle, lead to "blurring" of the energy dependence of the surface brightness profiles of the extended source, thus preventing the possibility of measurement of magnetic field via the study of the source extension. Besides, both spectrum and morphology of the extended emission could depend on the spectrum of the primary source. This dependence could lead to the loss of the characteristic $E_\gamma^{-1}$ behaviour of the source extension and to the loss of information about the properties of EGMF. 

In order to study these problems in more details, we have developed a Monte-Carlo code for the  modeling  of  \gr -induced electromagnetic cascades in the intergalactic space, with account of the EGMF.

%%%%%%%%%%%%%%%
\subsection{Primary \gr s}
%%%%%%%%%%%%%%%%%%%%%%%%

We consider, as a reference case, an extragalactic source that injects \gr s with a powerlaw energy distribution, $dN_{\gamma_0}/dE_{\gamma_0}\sim E_{\gamma_0}^{-\Gamma}\exp(-E_{\gamma_0}/E_{\rm cut})$ with the powerlaw index $\Gamma=2$ and cut-off energy $E_{\rm cut}=300$~TeV. The calculations presented below can be generalized (repeated) for an arbitrary assumptions about the primary source spectrum. As it is explained below, an analysis of the real data sets aimed at the measurement of EGMF has to include Monte-Carlo simulations of the properties of the extended sources produced by the primary point sources with different spectra.

To calculate the pair production by the primary \gr s in the intergalactic space, we assume the  energy dependence of the mean free path of the \gr s  calculated in the Ref. \cite{aharon} for "nominal" Cosmic-Infrared-Background (CIB) model. We take into account the uncertainty of the measurements of CIB by allowing for different overall normalizations of the CIB.  The pair production on the CMB becomes important at the energies above $\sim 50$~TeV. Since our calculations extend to the primary \gr s with energies above $50$~TeV we also take into account the CMB photon background. 

If the mean free path of the \gr s of the energy $E_{\gamma_0}$ is $D_{\gamma_0}$, the probability density of creation $e^{+}e^{-}$ pair at a distance $x$ from source is
\begin{equation}
\label{puas}
P(x)=\frac{1}{D_{\gamma_0 }}\int_0^{x}\exp\left(-\frac{x'}{D_{\gamma_0 }}\right)dx'.
\end{equation}
so that the distances $x_i$ at which $i$-th \gr\ creates an $e^+e^-$ pair could be expressed through a random number $P_i$  distributed in the interval $0<P_i<1$ as 
\begin{equation}
\label{rnd}
x_{i}=-D_{\gamma_0 }\cdot \ln(P_{i}) .
\end{equation}

Using the production spectrum of the $e^+e^-$ pairs, $dN(E_e)/dE_e=f(\omega _0, E_{\gamma_0}, E_e)$, where $\omega_0$ is the energy of the background photon, we calculate the energy of electron, produced by the $i$-th primary \gr, taking into account that the probability for an electron to have  energy $E_{e, i}$ is
\begin{equation}
\label{el}
P(E_{e,i})=\int_0^{E_{e,i}}f(\omega_0, E_{\gamma 0,i},E_e)dE_e
\end{equation}
Having calculated the electron energy $E_{e,i}$ we calculate the positron energy as $E_{e^{+},i}=E_{\gamma_0} -E_{e,i}$. Initial directions of motion of electron and positron ${\bf v}_{e^\pm}$ coincide with the primary photon direction. 

%%%%%%%%%%%%%%%%
\subsection{$e^+e^-$ pairs}
%%%%%%%%%%%%%%%%%

We model the propagation of both electrons and positrons in the EGMF by solving the equations of motion,
\begin{equation}
\label{mot}
\gamma_e m_{e}\frac{d{\bf v}_e}{dt}=\pm e\ {\bf v}_e\times {\bf B}
\end{equation}
here $\gamma_e$ is Lorentz factor of $e^{-}$ or $e^{+}$, $m_e,e$ are, respectively, electron mass and charge and ${\bf v}_e$ is its velocity. 

The IC scattering of the background photons is treated in a way similar to the one used for the pair production, via calculation of the distance of emission of each next IC \gr, using equation similar to (\ref{rnd}) with a substitution $D_{\rm IC}=(\sigma_T n_{\rm CMB})^{-1}$, where $\sigma_T$ is the Thomson cross-section, at the place of $D_{\gamma_0}$. The energy of each subsequent upscattered photon is calculated based on the differential cross-section of IC scattering \cite{blumenthal}, in a way similar to the calculation of the energies of the $e^+e^-$ pairs, described above. Having calculated the energy of an IC \gr, $E_\gamma$, we decrease the electron/positron energy, $E_e\rightarrow E_e-E_{\gamma }$. We repeat the generation of the IC \gr s  for all branches of the cascade until energies of electrons/positrons become less then 0.1 TeV. 

The secondary cascade \gr s produced via IC scattering process by the $e^+e^-$ paris could themselves be absorbed by interactions with CIB photons and inject additional $e^+e^-$ pairs. We take this possibility into account in our calculations.  We do not include the cosmological evolution of the CMB and CIB, because we consider only the relatively nearby extragalactic \gr\ sources, situated at the distances $\sim 100$~Mpc.

%%%%%%%%%%%%%%%%%%%%%%%%%%%%%%%%%
\subsection{Extragalactic Magnetic Field.} 
%%%%%%%%%%%%%%%%%%%%%%%%%%%%%%%%%

The energy attenuation length of the $e^+e^-$ pairs in the intergalactic medium is short compared to the coherence length of the EGMF, $D_{\rm EGMF}\sim 1$ Mpc, estimated to be of the order of the typical distance between the galaxies \cite{beck08,Harari}. This means that one can assume that electrons and positrons propagate in an ordered magnetic field (rather than diffuse in a random field) during their radiative cooling. In our modeling we divide all the space into grid of cubic cells with the side length 1 Mpc. In each cell we choose a constant magnetic field with random orientation. We assume, for simplicity, that the strength of EGMF is the same in each cell and leave the investigation of the effect of variations of the EGMF strength along the lines of sight toward individual extragalactic sources for the future work.

%%%%%%%%%%%%%%%%%%%%%%%%%%%%%%%%%%%%%%%%%%%%%%%%%%%
\subsection{Properties of \gr -induced  cascades in intergalactic space}
%%%%%%%%%%%%%%%%%%%%%%%%%%%%%%%%%%%%%%%%%%%%%%%%%%%

\begin{figure}
\includegraphics[width=1\columnwidth]{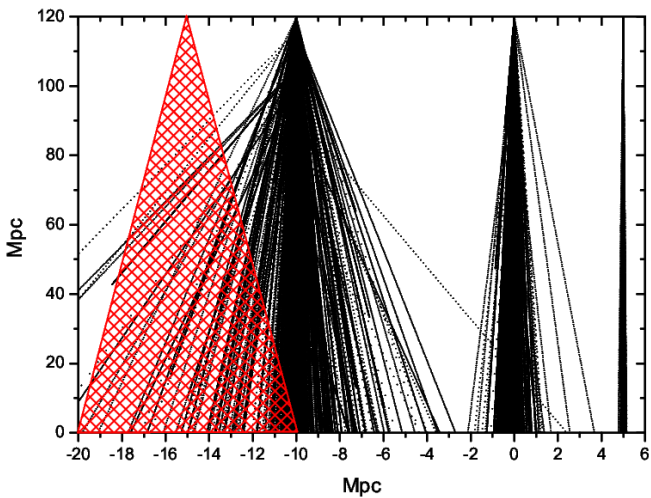}
\caption{Examples of showers from $\gamma $-rays with primary energy
$E_{0} = 300$ TeV developing in the EGMF of the strength $10^{-14}$ (left), $10^{-15}$ (middle) and $10^{-16}$ G (right). Only photon tracks with energies above  0.1 TeV are shown. The mean free paths of the electrons and positrons are much shorter ($\sim 1$ kpc) than the cascade development scale and they are not visible in the Figure.}
\label{fig:mf}
\end{figure}

\begin{figure}
\includegraphics[width=1\columnwidth]{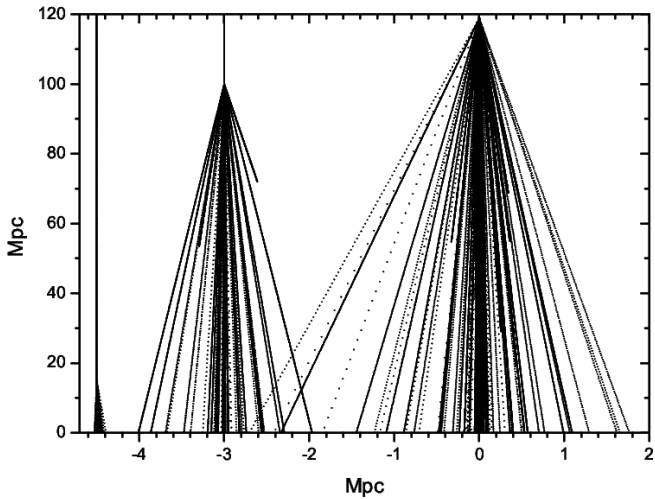}
\caption{Examples of showers from $\gamma $-photons with different primary energies 10 (left), 30 (middle) and 100 TeV (right) in EGMF $10^{-15}$ G.}
\label{fig:E0}
\end{figure}

The results of typical runs of our code are presented in the  Figs. \ref{fig:mf} and \ref{fig:E0}. Fig. \ref{fig:mf} compares the cascades induced by the \gr s of the energy $E_{\gamma_0}=300$~TeV emitted from a source at the distance $D=120$~Mpc. The three cases shown with black color in the figure correspond to different choices of the strength of EGMF. For comparison, we show in the same figure in red a cone with an opening angle $\Theta_{\rm jet}=5^\circ$ which is believed to be a typical opening angle of the blazar jets ($\Theta_{\rm jet}\simeq 1/\Gamma_{\rm jet}$, where $\Gamma_{\rm jet}\sim 10$ is the bulk Lorentz factor of the jet). The leftmost black-colored  cascade develops in the EGMF of the strength $B=10^{-14}$~G, the cascade in the middle develops in EGMF $B=10^{-15}$~G while the cascade on the right develops in a magnetic field $B=10^{-16}$~G. One can clearly see that the increase of the strength of the magnetic field leads to the increase of the opening angle of the cascade. 
Also the opening angle of the cascade becomes larger than the typical jet opening angle at a certain distance from the source. This distance depends on the EGMF strength. 

Fig. \ref{fig:E0} shows the dependence of the geometrical properties of the cascade on the energy of the primary \gr. The left, middle, and right cascades are initiated by the \gr s of the energies $10, 30$ and $100$~TeV, respectively. The decrease of the mean free path of the primary \gr\  with the increase of energy is evident from this figure. The decrease of the mean free path leads to the earlier on-set of the electromagnetic cascade. 

%%%%%%%%%%%%%%%%%%%%%%%%%%%%%%%%%%%%%%%%%%%%%%%%%%%
%%%%%%%%%%%%%%%%%%%%%%%%%%%%%%%%%%%%%%%%%%%%%%%%%%%
%%%%%%%%%%%%%%%%%%%%%%%%%%%%%%%%%%%%%%%%%%%%%%%%%%%
%%%%%%%%%%%%%%%%%%%%%%%%%%%%%%%%%%%%%%%%%%%%%%%%%%%

%%%%%%%%%%%%%%%%%%%%%%%%%%%%%%%%%%%%%%%%%%%%%%%%%%%
\subsection{Extended emission around an extragalactic point source}
%%%%%%%%%%%%%%%%%%%%%%%%%%%%%%%%%%%%%%%%%%%%%%%%%%%

\begin{figure}
\includegraphics[width=0.9\columnwidth]{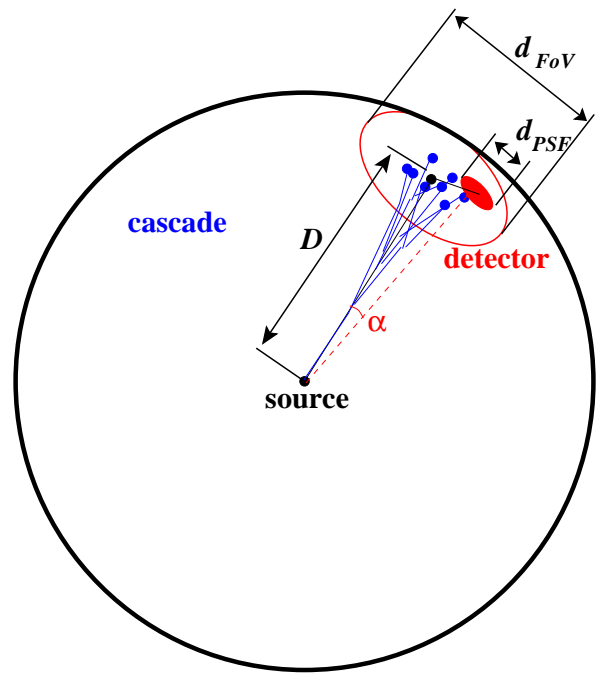}
\label{fig:e0}
\caption{Scheme of Monte-Carlo calculations of the extended emission.}
\label{fig:MC_scheme}
\end{figure}

To model the detection of the \gr -induced cascades by the space- or ground-based \gr\ telescope, we adopt the following procedure, illustrated by Fig. \ref{fig:MC_scheme}. 

Simulations of electromagnetic cascades initiated by the primary \gr s injected by an isotropically emitting source  or by a source emitting into a jet with a certain opening angle take a lot of computing time. Besides, most of the cascade photons initially emitted at large angles with respect to the line of sight would never reach the detector, so that the machine time spent on their calculation would be wasted. Taking this fact into account, we  inject all the primary photons in the same direction. We save the coordinates and arrival directions of all the cascade photons with energies above 0.1~TeV, which crossed the sphere with centre at the location of the source and with radius $D$ equal to the distance to the detector. 

%%%%%%%%%%%%%%%%%%%%%%%%%%%%%%%%%%%%%%%%%%%%%%%%%%%
%%%%%%%%%%%%%%%%%%%%%%%%%%%%%%%%%%%%%%%%%%%%%%%%%%%

Next, to take into account the characteristics of the detector, we choose a circle of the angular radius $d_{\rm PSF}=D\theta_{\rm PSF}$ where $\theta_{\rm PSF}$ is the angular resolution of the detector. To calculate the extended emission produced by an isotropically emitting source,  we place the detector circle at a set of random positions on the sphere and sum all the cascade photons which pass through the detector circle at a given angle with respect to the normal to the sphere.  This procedure is equivalent to injection of primary photons in different directions at fixed detector but is more time-efficient.  For the sake of simplicity, we restrict our calculations here to the case when the source emits \gr s isotropically. In this case the effects of distortion of the circular symmetry of the surface brightness of the extended source due to the finite opening angle of the blazar jet and misalignment of the jet with respect to the line of sight do not affect the measurement of EGMF strength. We leave the investigation of the distortions of the morphology of extended emission caused by specific geometries of the jet for the future investigation.

\begin{figure}
\includegraphics[width=1\columnwidth]{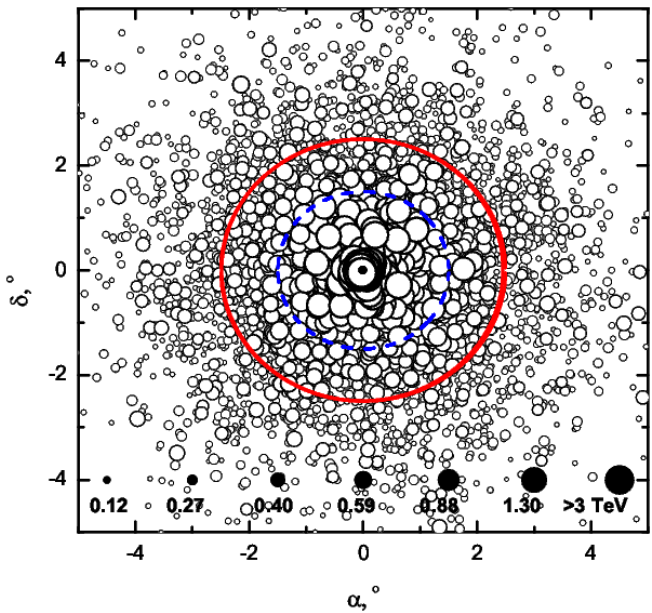}
\caption{The arrival directions of the primary and secondary cascade \gr s (circles) from a source
at a distance $D =120$~Mpc.  The EGMF strength is $10^{-14}$~G. The sizes of the circles representing each photon are proportional to the photon energies. The blue dashed circle has radius $1.5^\circ$, equal to the radius of the FoV or MAGIC telescope. The radius of the blue solid circle is $2.5^\circ$, which corresponds to the size of the FoV of HESS telescope.}
\label{fig:halo-14}
\end{figure}

\begin{figure}
\includegraphics[width=1\columnwidth]{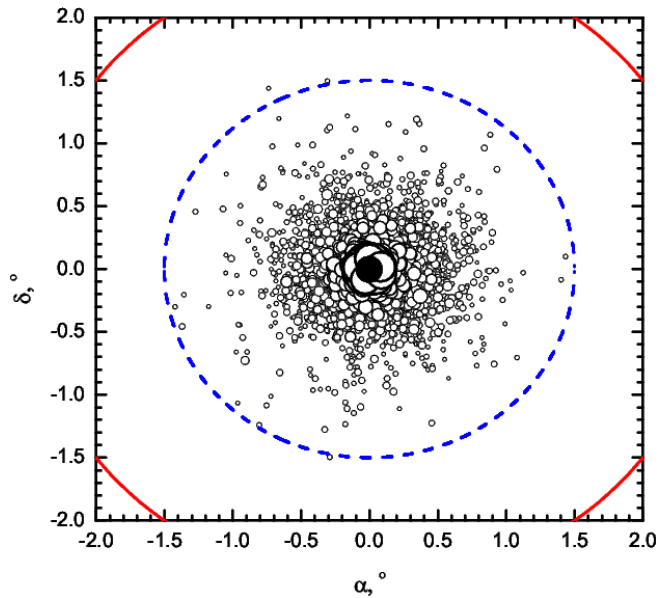}
\caption{Same as Fig. \ref{fig:halo-14} but for the EGMF strength $10^{-15}$G.}
\label{fig:halo-15}
\end{figure}

The results of numerical calculation of the energy-dependent morphology of extended emission are shown in Figs. \ref{fig:halo-14} and \ref{fig:halo-15} for a source at a distance $D=120$~Mpc (equal to the distance of the blazar Mrk 421).

 In the case of calculation shown in Fig. \ref{fig:halo-14} the EGMF strength is assumed to be $B=10^{-14}$~G.  One can see that strong deflections of the cascade electrons by the EGMF lead in this case to a very large size of the extended emission at the energies $\sim 0.1$~TeV, with significant contribution to the signal extending up to $\sim 4^\circ$ from the source. Obviously, the extended emission at these energies can not be detected by a typical Cherenkov telescope array, which has a FoV with the diameter   $\sim 5^\circ$ (HESS) or $\sim 3^\circ$ (MAGIC) (shown as, respectively, solid and dashed circles in Fig. \ref{fig:halo-14}). The extension of the source becomes smaller than the size of HESS and MAGIC FoVs only at the energies above $\sim 1$~TeV.

%%%%%%%%%%%%%%%%%%%%%%%%%%%%%%%%%%%%%%%%%%%%%%%%%%%
%%%%%%%%%%%%%%%%%%%%%%%%%%%%%%%%%%%%%%%%%%%%%%%%%%%

%%%%%%%%%%%%%%%%%%%%%%%%%%%%%%%%%%%%%%%%%%%%%%%%%%%
%%%%%%%%%%%%%%%%%%%%%%%%%%%%%%%%%%%%%%%%%%%%%%%%%%%

The extended emission is more compact if the strength of EGMF is weaker. This is illustrated in Fig. \ref{fig:halo-15} which shows the result of calculation of extended emission around the same point source as in Fig. \ref{fig:halo-14}, but assuming the EGMF strength $B=10^{-15}$~G. One can see that in this case most of the extended emission down to the energies $\sim 0.1$~TeV fits into the FoVs of both HESS and MAGIC telescopes. 
 
\begin{figure}
\includegraphics[width=1\columnwidth]{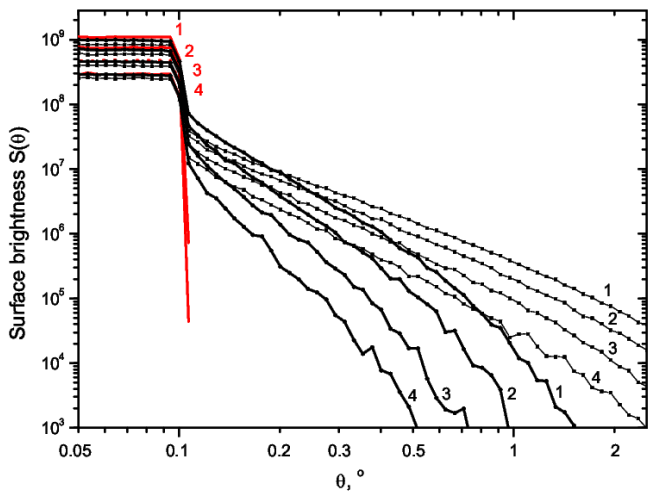}
\caption{Surface brightness profiles of emission in different energy bands (1 -- 0.10-0.12~ TeV, 2 -- 0.12-0.18~TeV, 3 -- 0.18-0.27~TeV, 4 -- 0.27-0.40 TeV) produced by the cascades in EGMF of the strength $10^{-14}$G (thin lines) and $10^{-15}$G EGMF (thick lines) and without magnetic field (red lines).}
\label{fig:profile}
\end{figure}

Fig. \ref{fig:profile} shows the radial surface brightness profiles of the extended sources shown in Figs. \ref{fig:halo-14} and \ref{fig:halo-15} in different energy bands. One can see that in each energy band the profiles of the extended emission produced by the cascade in a weaker magnetic field (open diamonds) are always more concentrated toward the central source than the profiles of extended emission produced by a cascade in a stronger magnetic field (filled diamonds, the magnetic field is $10^{-14}$~G. 
The details of the shape of the surface brightness profile at the angular scale $\theta\sim 0.1^\circ\sim \theta_{\rm PSF}$ close to size of the PSF of a telescope could be calculated only if the details of the response of the telescope are known. 

In our calculations we do not model the telescope's PSF and, instead, assume that \gr s from a point source are homogeneously distributed within a disk of the radius $\theta_{\rm PSF}=0.1^\circ$. This explains the appearance of "plateau" in the surface brightness profiles. Comparing the normalization of the plateau with the one of the extended emission, one can judge, if the sensitivity of a telescope is sufficient for the detection of extended emission around a point source at a given level of the significance of detection of the source for a particular telescope. It is important to note that, in spite of the much lower normalization of the surface brightness profiles of extended emission, as compared to the point source emission, the total flux of the extended source (integrated over the telescope's field of view) could be comparable to the one of the point source.

%%%%%%%%%%%%%%%%%%%%%%%%%%%%%%%%%%%%%%%%%%%%%%%%%%%
%%%%%%%%%%%%%%%%%%%%%%%%%%%%%%%%%%%%%%%%%%%%%%%%%%%

The qualitative discussion of the model radial brightness profile of the extended emission, presented in the previous section, adopted simplifying assumptions (such as e.g. the monochromatic spectrum of the primary \gr\ source and monochromatic approximation for the spectrum of IC emission from the cascade $e^+e^-$ pairs). Although useful for the qualitative understanding of the method of measurement of EGMF, these assumptions do not hold in realistic situation. In a more general case, the shape of the radial surface brightness profile of extended source, $S(\theta,E)$, could be derived from the results of our numerical calculations. 

In the particular case of a powerlaw primary \gr\ spectrum with the photon index $\Gamma=2$, which we consider as an example, the radial brightness profiles of extended source shown in Fig. \ref{fig:profile} can be approximated as  powerlaw with an exponential cutoff:
\begin{equation}
\label{fun}
\frac{dn_\gamma (E_\gamma)}{d\theta}\propto \theta S(\theta,E_\gamma)\propto \left(\frac {\theta}{\Theta_{\rm cut}(E_\gamma)}\right)^{\alpha }\exp\left(-\frac{\theta}{\Theta_{\rm cut}(E_\gamma)}\right),
\end{equation}
with the slope $\alpha\simeq -1$. The energy-dependent parameter $\Theta _{\rm cut}(E_\gamma)$ characterizes the extension of the source at a given energy $E$.  This parameter is an analog of the parameter $\Theta_{\rm ext}$, which would characterize the size of the extended source in the case of the monochromatic primary source spectrum. 

%%%%%%%%%%%%%%%%%%%%%%%%%%%%%%%%%%%%%%%%%%%%%%%%%%
\subsection{Measurement of EGMF strength}
%%%%%%%%%%%%%%%%%%%%%%%%%%%%%%%%%%%%%%%%%%%%%%%%%%

\begin{figure}
\includegraphics[width=1\columnwidth]{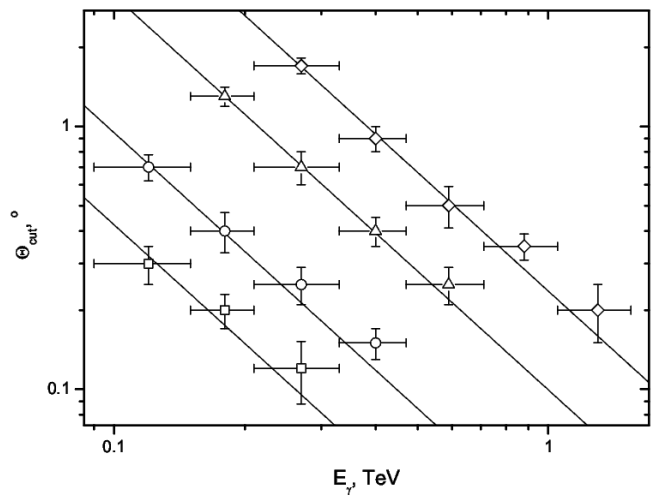}
\caption{$\Theta _{\rm cut}$ as function of energy $E_{\gamma }$ for several values of  EGMF, $10^{-15}$ G (squares), $2\cdot 10^{-15}$ (circles), $5\cdot 10^{-15}$ (triangles), $10^{-14}$ (diamonds). Solid lines correspond to the fits of the data points with $\Theta_{\rm cut}\sim E_\gamma^{-1}$.}
\label{fig:teta0}
\end{figure}

Since the parameter $\Theta_{\rm cut}$, which appears in Eq. (\ref{fun}), is a direct analog of the parameter $\Theta_{\rm ext}$ for the case of a source emitting a powerlaw spectrum of primary \gr s, one expects that, similarly to $\Theta_{\rm ext}$, $\Theta_{\rm cut}$ should decrease inversely proportionally to the  energy. Fig. \ref{fig:teta0}
shows that this is indeed the case. 

The energy dependence of the size of the extended source, $\Theta_{\rm cut}(E_\gamma)$ is determined by the strength of EGMF along the line of sight. Therefore, measuring this energy dependence, one can extract information about the properties of EGMF from the observational data. In the simplest case of a monochromatic point source of \gr s, the dependence of the source extension on the EGMF strength could be found analytically (\ref{Thetaext}). It is clear that  if the primary \gr s from the source are distributed within a certain energy range, a superposition of the extended sources produced in result of absorption of the primary \gr s of different energies will lead to the changes of the normalization coefficient $N=N(\Gamma,E_{\rm cut})$  depending on the spectral model of the primary source. Then  Eq. (\ref{Thetaext}) can be rewritten for the case of power law injection spectrum via substitution $\Theta_{\rm ext}\rightarrow \Theta_{\rm cut}$ and introduction of $N(\Gamma,E_{\rm cut})$:
\begin{equation}
\label{Thetacut}
\Theta_{\rm cut} (E_\gamma,B)= \frac{N(\Gamma,E_{\rm cut})}{\tau\left(E_{\gamma_0}(E_\gamma),z\right)}\left[\frac{E_\gamma}{6\mbox{ TeV}}\right]^{-1}\left[\frac{B}{10^{-13}\mbox{ G}}\right]~.
\end{equation}
Using a numerically calculated $N(\Gamma,E_{\rm cut})$, one would be able to derive the strength of EGMF from the observationally measured $\Theta_{\rm cut}(E_\gamma)$.  

To illustrate the implementation of the algorithm described above,  we consider the case of a primary \gr\ source with the photon index $\Gamma=2$ and and exponential high-energy cut-off at $E_{\rm cut}=300$~TeV. We derive the value of the normalization coefficient $N(\Gamma=2,E_{\rm cut}=300\mbox{ TeV})$ from a fit of the energy dependence of $\Theta_{\rm cut} (E_\gamma)$ using the simulated properties of the extended emission for a (arbitrarily chosen) reference EGMF $B= 10^{-15}$~G. Once the value of the normalization coefficient $N$ is fixed, one can resolve the Eq. (\ref{Thetacut}) with respect to $B$ and use the measurements of $\Theta_{\rm cut}(E_\gamma)$ to infer the value of $B$ from the simulated data sets produced for different choices of  EGMF strength. 

\begin{figure}
\includegraphics[width=1\columnwidth]{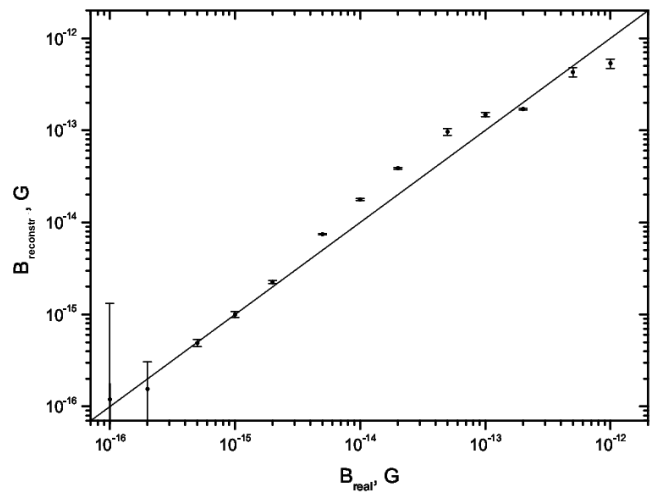}
\caption{ Comparison of the EGMF strength reconstructed from measurement of extended emission in simulated data sets, $B_{\rm reconstr}$, with the real value of the magnetic field strength assumed for the simulations, $B_{\rm real}$. The line corresponds to $B_{\rm real}=B_{\rm reconstr}$}
\label{fig:res}
\end{figure}

The result of such a procedure of "reconstruction" of $B$ from the simulated data sets is shown in  Fig. \ref{fig:res}. The $y$ axis of the plot shows the reconstructed value of the EGMF strength, while the initially assumed EGMF strength is plotted along the $x$ axis. Since the normalization coefficient $N$ is found from the simulations assuming the "reference" magnetic field, the reconstructed magnetic field is exactly equal to the assumed magnetic field at the reference point $B=10^{-15}$~G.
 
From Fig. \ref{fig:res} one can see that the assumed value of EGMF can be reconstructed from the data in a satisfactory way only in a certain range of EGMF strengths ($10^{-16}\mbox{ G}<B<10^{-12}\mbox{ G}$). This range is determined by the assumed characteristics of the \gr\ telescopes and/or by the intrinsic properties of the cascade. 

When the field strength gets close to $10^{-16}$~G, the size of the extended source at the energies above   $\sim 0.1$~TeV approaches the assumed angular resolution of the telescope, $\theta_{\rm PSF}=0.1^\circ$.  Under these conditions, the extended emission becomes indistinguishable from the point source and measurement of $\Theta_{\rm cut}(E)$ becomes impossible. This is reflected in a large errorbar of the point at $10^{-16}$~G in Fig. \ref{fig:res}. Obvious ways to extend the range of accessible EGMF strengths toward still weaker fields is to improve the angular resolution and/or to extend the observed energy band to the energies below 0.1~TeV.  Another possibility is to observe such field with {\it Fermi} at the  energies 1-100 GeV.

If the EGMF along the line of sight is stronger than $10^{-14}$~G, the size of extended source at the energy $\sim 0.1$~TeV becomes comparable to the size of the FoV of the current generation Cherenkov telescope arrays, $\theta_{\rm FoV}\sim 2^\circ$. Under these conditions, the measurement of $\Theta_{\rm cut}(E)$ is still partially possible at the energies much larger than $\sim 0.1$~TeV, but the precision of determination of $\Theta_{\rm cut}(E)$ rapidly drops when the field becomes larger than $\sim 10^{-14}$~G. One should note that, in a realistic situation, the measurement of the radial brightness profile of extended source becomes problematic already when the source size is smaller, but comparable to the size of the field of view. The difficulty is that the source signal has to be disentangled from the cosmic ray background, in the camera, with intensity depends on the off-axis angle.  

In our calculations we have ignored these difficulties and assumed that (a) the telescope FoV is not limited and (b) that an arbitrary large source (with extension up to $\sim 10^\circ$) could be detected. One can see that with a large FoV, high sensitivity telescope the field strength can be reconstructed from the measurement of characteristics of the extended emission up to the EGMF strengths $\sim 10^{-12}$~G, at which the energy dependence of the source size on EGMF gets lost because of the too strong deflections of the cascade electrons and positrons.

In principle, an energy-dependent size of the extended source is predicted also when the strength of EGMF is larger than $10^{-12}$~G, within the model of extended emission discussed in the Ref. \cite{Aharonian:halo}. However, in this case the size of the source is determined by the mean free path of the 10-100~TeV \gr s through the cosmological infrared/optical background. The angular extension of the source in this model can be readily estimated to be $\Theta(E_\gamma)\sim \left[\tau(E_{\gamma_0})\right]^{-1}$. Taking into account the fact that in a limited energy range $\tau(E_{\gamma_0})$ could be approximated by a linear function, one finds that the size of the extended source is also expected to shrink with energy. Contrary to the case of the EGMF-dependent source size, the size of the halos considered in the Ref. \cite{Aharonian:halo} decreases as $\Theta(E_\gamma)\sim E_{\gamma_0}^{-1}\sim E_\gamma^{-1/2}$ (see Eq. (\ref{Esec})). This means that the energy-dependence of the size of extended halos around extragalactic sources considered in the Ref. \cite{Aharonian:halo} could be readily distinguished from the one expected if the extended source size is determined by the EGMF. Systematic detection of the extended emission with the characteristic $E_\gamma^{-1/2}$ dependence of the source extension on the energy would provide a strong argument in favor of the existence of strong EGMFs, $B\gg 10^{-12}$~G.

At the end of this section, we would like to comment on the implementation of the proposed method for the analysis of the data of the real \gr\ observations of extragalactic sources. In a realistic situation the primary source spectrum is (a) not known in advance and (b) is strongly modified at the highest energies by the effect of absorption on the EBL. This means that, in general, one can not assume a particular spectral model of the primary sources for the EGMF search analysis, as it is done above for the analysis of the Monte-Carlo simulated data sets. Instead, the measurement of EGMF has to be done simultaneously with the measurement of the parameters of the high-energy emission spectrum of the primary source. This implies the following analysis procedure. First, using the Monte-Carlo simulations, described above, one has to produce a "look-up table" of the normalization coefficients $N(\Gamma,E_{\rm cut},...)$ for a set of parameters $(\Gamma, E_{\rm cut},....)$ of different possible spectral models of  the \gr\ source. Next, using these look-up tables, one can simultaneously find the best-fit values for the spectral model parameters $(\Gamma, E_{\rm cut},...)$ and for the EGMF strength $B$, via a multi-parameter fitting the spectral and morphological characteristics of the extended emission. 

%%%%%%%%%%%%%%%%%%%%%%%%%%%%%%%%%%%%%%%%%%%%%%%%%%%%%%%%%%%%%%%%%%%%%%%%%%
%%%%%%%%%%%%%%%%%%%%%%%%%%%%%%%%%%%%%%%%%%%%%%%%%%%%%%%%%%%%%%%%%%%%%%%%%%

%%%%%%%%%%%%%%%%%%%%%%%%%%%%%%%%%%%%%%%%%%%%%%%%%%%%%%%%%%%%%%%%%%%%%%%%%%
%%%%%%%%%%%%%%%%%%%%%%%%%%%%%%%%%%%%%%%%%%%%%%%%%%%%%%%%%%%%%%%%%%%%%%%%%%

%%%%%%%%%%%%%%%%%%%%%%%%%%%%%%%%%%%%%%%%%%%%%%%%%
\section{Summary and Conclusions}
\label{sec:conc}
%%%%%%%%%%%%%%%%%%%%%%%%%%%%%%%%%%%%%%%%%%%%%%%%%

In this paper we have used Monte-Carlo simulations of development of \gr\ induced electromagnetic cascade in the intergalactic space to study the properties of extended emission around extragalactic point sources of VHE \gr s. We have demonstrated that the properties of the cascade and, as a consequence, the properties of the expected extended emission around extragalactic \gr\ sources  strongly depend on the strength of the (largely uncertain) EGMF along the line of sight toward the source. 

The dependence of the properties of the cascades on the  EGMF strength could be used to reveal the presence of extremely weak magnetic fields in the intergalactic medium and to measure (or, at least, estimate) the strength of these fields. Such a measurement would enable to distinguish between existing competing cosmological models of the origin of magnetic fields in the Universe, which widely differ (by more than 10 orders of magnitude) in their predictions of the EGMF strength.

We have demonstrated that if the typical magnetic field strength along the line of sight is in the range $10^{-16}$~G$\le B\le 10^{-12}$~G, the cascade emission could be detected as an energy-dependent extended "glow" around the point source (see Figs. \ref{fig:halo-14}, \ref{fig:halo-15}) by the ground-based \gr\ telescopes. Telescopes optimized for the searches of stronger and weaker magnetic fields should have quite different characteristics. Large effective collection area and large size of the telescope's field of view are needed for the detection of stronger magnetic fields $B\sim 10^{-12}$~G.  At the same time, high sensitivity at low \gr\ energies (below 0.1~TeV) and good angular resolution are important for the detection of weaker $B\sim 10^{-16}$~G fields. Still weaker fields $B\le 10^{-16}$~G should reveal themselves through the extended emission around extragalactic \gr\ sources in the {\it  Fermi} energy band, $E_\gamma \sim 1-100$~GeV.

The extension of the \gr\ glow around extragalactic point sources appears to decrease inversely proportionally to the energy (Figs. \ref{fig:profile}, 
\ref{fig:teta0}). Measurement of the energy dependence of the source size enables to derive a constraint on the strength of extragalactic magnetic field from the data (Monte-Carlo simulated data in our case), using Eq. (\ref{fun}), (\ref{Thetacut}). We have described the algorithm of reconstruction of the EGMF strength from the data.  Fig. \ref{fig:res} presents the results of implementation of the algorithm for the analysis of the simulated data sets. The proposed algorithm could be directly used for the analysis of the real data of ground and space- based \gr\ telescopes. 

Apart form the measurement of the EGMF, a study of the properties of extended emission around extragalactic point sources enables to constrain the properties of the primary \gr\ source spectrum at the energies above 10~TeV at which the flux from the source is strongly suppressed by absorption on the extragalactic background light.  We have argued that, in fact, in the analysis of the real data, the constraints on the spectral characteristics of the primary source have to be found simultaneously with the constraints on the EGMF strength.

When this paper was finished, we became aware of a paper which analyzes the properties of extended emission around extragalactic \gr\ sources by Dolag et al. \cite{dolag09}, which presents results complementary to our study. Contrary to our study, the Ref. \cite{dolag09} concentrates on the analysis of  the properties of the EGMF-dependent extended emission assuming different spectral and morphological characteristics of the primary \gr\ source, fixing a particular model of EGMF.

\section*{Acknowlegdement} AE and AN are  grateful to Ukrainian Virtual Roentgen and Gamma-Ray Observatory VIRGO.UA and computing clusters of Main Astronomical Observatory and Bogolyubov 
Institute for Theoretical Physics, for using their computing resources. This work was supported by the Swiss National Science Foundation and the Swiss Agency for Development and Cooperation in the framework of the program SCOPES - Scientific cooperation between Eastern Europe and Switzerland.

\end{document}